\documentclass[conference]{IEEEtran}
\IEEEoverridecommandlockouts
\usepackage{tikz}
\newcommand\copyrighttext{%
  \footnotesize C. Matthies, A. Treffer and M. Uflacker, ``\textit{Prof. CI: Employing continuous integration services and Github workflows to teach test-driven development},'' 2017 IEEE Frontiers in Education Conference (FIE), Indianapolis, IN, 2017, pp. 1-8. doi: \href{https://doi.org/10.1109/FIE.2017.8190589}{10.1109/FIE.2017.8190589} \\[4pt]
  Copyright \textcopyright 2017 IEEE. Personal use of this material is permitted.
  Permission from IEEE must be obtained for all other uses, in any current or future 
  media, including reprinting/republishing this material for advertising or promotional 
  purposes, creating new collective works, for resale or redistribution to servers or 
  lists, or reuse of any copyrighted component of this work in other works.
  }
\newcommand\copyrightnotice{%
    \begin{tikzpicture}[remember picture,overlay]
    \node[anchor=south,yshift=10pt] at (current page.south) {\fbox{\parbox{\dimexpr\textwidth-\fboxsep-\fboxrule\relax}{\copyrighttext}}};
    \end{tikzpicture}%
}

\usepackage{lmodern} % use font that looks identical but has less bugs
\usepackage{csquotes}
\MakeOuterQuote{"}

\usepackage{graphicx} % Allow \includegraphics
\usepackage{tabularx} % The more x, the better
\usepackage{url} % \url
\usepackage{listings} % source code highlighting

\usepackage{enumitem}
\setlist{nosep}

\usepackage[]{acro}

\DeclareAcronym{MOOC}{long=Massive Open Online Course}
\DeclareAcronym{CI}{long=continuous integration}
\DeclareAcronym{BDD}{long=behavior-driven design}
\DeclareAcronym{TDD}{long=test-driven development}
\DeclareAcronym{VCS}{long=version control system}
\newcommand{\MOOC}{\ac{MOOC}}
\newcommand{\MOOCs}{\acp{MOOC}}
\newcommand{\ci}{\ac{CI}}
\newcommand{\bdd}{\ac{BDD}}
\newcommand{\tdd}{\ac{TDD}}
\newcommand{\vcs}{\ac{VCS}}

\newcommand{\TDD}{\Ac{TDD}}

\usepackage{hyperref}
\usepackage[all]{hypcap}
\usepackage{cleveref}

\begin{document}
\title{Prof. CI: Employing Continuous Integration Services and Github Workflows to Teach Test-driven Development}

\author{\IEEEauthorblockN{Christoph Matthies, Arian Treffer, and Matthias Uflacker}
\IEEEauthorblockA{Hasso Plattner Institute, University of Potsdam\\
August-Bebel-Str. 88\\
14482 Potsdam, Germany\\
Email: \{firstname.lastname\}@hpi.de}}


\maketitle

% Render IEEE copyright notice (after \maketitle)
\copyrightnotice

\begin{abstract}
Teaching programming using \MOOCs\ is gaining popularity due to their scalability and efficiency of knowledge distribution.
However, participating in these courses usually means fully committing to the supplied programming environment in the browser.
While this allows a consistent and controllable setup, learners do not gain experience with actual development tools, such as local code editors, testing frameworks, issue trackers or \ci\ services, which is critical for subsequent real-world projects.
Furthermore, the tests for the functionality that is to be developed are oftentimes already available in \MOOCs\ and simply need to be executed, leading to less involvement with developing appropriate tests.
In order to tackle these issues while maintaining a high degree of automation and scalability, we developed Prof. CI, a novel approach to conducting online exercises. 
Prof. CI leverages the existing automation infrastructure that developers use daily, i.e. \ci\ services and Github workflows, to teach \tdd\ practices.
Participants work on their own repositories in Github and receive feedback and new challenges from the CI server when they push their code.
We have successfully applied this approach in a pilot project with 30 undergraduate students learning the Ruby on Rails web development framework.
Our evaluation shows that the exercise effectively increased students' motivation to write tests for their code.
We also present the results of participant surveys, students' experiences and teachers' observations.
\end{abstract}

% Actually include the content
\section{Introduction}
Online courses can efficiently teach a large number of students.
Automating the evaluation and grading of student work allows to provide quick feedback and enables scalability.
In computer science, online courses can be used to teach programming, languages, and frameworks, such as Ruby on Rails; or methodologies such as \bdd\ and \tdd~\cite{Vihavainen2012}.
Other advantages of online courses include reporting and analytics possibilities that help teachers to improve exercises and identify students in need of targeted help~\cite{Fournier2011}.

\subsection{Problem Statement}
In online courses, programming tasks often happen entirely in the browser.
Participants do not learn to use local program\-ming-related tools, such as their IDE, a \vcs, or \ci, all of which are vital for working effectively in practice.
Because of the granularity of coding tasks, they also often don't learn to work on a full project, e.g. where certain files, such as configurations, are located, or how to test and build the application.
When teaching methodologies, a balance has to be found between the structure of the exercise and individual freedom.
With too much structure, students follow tasks without reflecting their purpose, with too much freedom, they may stop correctly using the methodology.
For these reasons, when we used online courses to prepare students for a software project, the participants reported that the learnings were too theoretical to be used in practice and more preparation time was needed.

\subsection{Research Questions}

The following research questions guided our research:
\begin{itemize}
    \item Can we lead students towards the habit of writing tests in software projects using introductory programming exercises?
    \item Can these exercises be implemented in a highly automated, yet student-centered manner?
\end{itemize}

\subsection{Approach}

In this paper, we present \emph{Prof. CI}, a new form of programming exercise to teach programming skills and \tdd.
Students work on their local machines, and a \vcs\ and \ci\ server are used to evaluate tasks and provide feedback.
However, unlike in traditional programming exercises where the entire assignment is specified up-front, Prof. CI reveals requirements in small tasks as tickets in an issue tracker.
This happens automatically based on the student's progress and is designed to reinforce the workflows of \tdd.

This approach combines the benefits of both online and conventional programming exercises.
Compared to online courses, students gain more practical experience by immediately working on a full project and using actual development tools.
Nevertheless, students are guided in small steps of increasing difficulty and receive immediate feedback on their work.
Furthermore, the tight interaction with the exercise systems allows us to influence how students approach the coding assignments, in particular regarding the test-first rule of \tdd.

Our evaluation shows that a Prof. CI exercise positively influences student behavior in subsequent software projects, compared to online courses and conventional exercises.

The evaluation data, as well as all source code and additional documentation on our prototype exercise is publicly available\footnote{\url{https://hpi-epic.github.io/profci-exercise/}}.
The exercise runs fully on free online services such as GitHub and Travis CI, which gets students familiarized with existing popular technologies.
However, it is also possible to use private, self-hosted infrastructure, which might be preferable in some cases.

The remainder of this paper is structured as follows:
\Cref{sec:related-work} discusses related work. 
\Cref{sec:prof-ci} presents the details of our approach.
\Cref{sec:evaluation} reports how we used Prof. CI in practice and evaluates the effectiveness of our approach. 
\Cref{sec:conclusion} concludes.

\section{Related Work}
\label{sec:related-work}
Jones~\cite{jones2004test} summarizes four experience reports, one conceptual paper, and three experiments concerning pedagogical experiences with using a test-first approach in the classroom.
While the author points out that evaluation of the purported benefits of \tdd\ yielded only  mixed results, he concludes that \tdd\ shows promise as a means to help students achieve a verifiable design specification.

In a later study, Desai et al.~\cite{desai2008survey} conducted a survey of \tdd\ usage in academia, analysing 18 case studies and experience reports.
They highlight the positive effects of \TDD, including increased student confidence.
However, the authors also note that the best learning outcomes were achieved by more mature students of junior undergraduate and graduate levels.
This is encouraging as we employed Prof. CI with undergraduate students with prior programming experience.

Janzen and Saiedian~\cite{janzen2006test} introduced Test-driven learning (TDL), a
pedagogical tool which involves introducing new programming concepts through complete unit tests.
By including assertions together with examples, the use of interfaces as well as the expected behavior of the software element is documented.
The authors list "encouraging the use of TDD" as one of TDL's goals.
TDL highlights the importance of testing in education.

Spacco and Pugh~\cite{spacco2006helping} introduced \emph{Marmoset}, a system for student programming project submission and testing.
It lays a strong focus on increasing student motivation concerning the writing of tests.
Projects including skeleton starter code and initial test cases are distributed to students.
The authors point out that \tdd\ heavily relies on rapid feedback, so students can submit their code at any time to Marmoset, where it is tested using the student's tests, in a traditional \ci\ setup.
Additionally, students can request a \emph{release test}, which triggers a set of secret, teacher supplied, tests.
Students are supplied the names of failing release tests.
The process of release testing consumes a \emph{token}, which is limited in supply.
The goal is to encourage students to write their own test cases and motivate them to start doing so earlier, so that more opportunities for release testing are available.
Marmoset presents an interesting approach for increasing the need for student testing.
However, some caveats apply, such as the need to use custom software that students need to be familiarized with.
Furthermore, communicating only the names of failing test cases might not be enough information to write meaningful test cases and makes it hard for educators to provide additional helpful resources.

Staubitz et al.~\cite{staubitz2015towards} provide an overview of automated assessment approaches with a focus on MOOCs. 
While they argue that beginners benefit more from a browser-based environment, which relieves them from ``the agony of installation hassles'', they also point out more advanced users.
They are more inclined to stick with familiar tools they have already installed.
The authors, furthermore conclude, that a ``code local --- assess remote'' approach has benefits, including not having to provide an immediate response.

Similarly, Fox et al.~\cite{fox2015magic} employed Continuous Integration services in conjunction with a software engineering MOOC as well as in Small Private Online Courses (SPOCs).
Their \emph{MAGIC} approach uses \ci\ tasks to automatically install and run an autograder software.
When a student creates a new pull request, the autograder checks the code against a set of known solutions and gives feedback to the student.
Much like in our courses, the authors point out that the exercises are used as formative rather than summative assessments.
As such, the issue of cheating is not explicitly addressed by the software.

Others have used \ci\ services in education mostly for their original use.
Billingsley et al.~\cite{billingsley2013comparison} employed \ci\ practices in a software engineering course with 70 participants working on real-world legacy code.
They point out that this allowed them to use a realistically sized project without needing 
significant extra staffing.

We aim to employ standard \ci\ services to provide detailed error reports or user stories to be worked on next to students using the Prof. CI system.
Rather than having the CI system itself act as the feedback mechanism, GitHub issues containing these details are created in the student's repository when educator tests are not passed.

\section{Professor CI}
\label{sec:prof-ci}
\acresetall

\emph{Prof. CI} is a new approach for teaching software development using \ci\ services to automatically manage participants' work.
It allows students to work on their local machine and gain experience with the tools and processes they will apply in subsequent real-world software development projects.
Participants have the freedom to decide how to implement a requirement, while also being encouraged to use agile methods, such as \tdd.
As in a web-based programming exercise, Prof. CI allows the automatic evaluation and detailed reporting of participants' work.

\subsection{Exercise Procedure}

\begin{figure}[t]
    \centering
    \includegraphics[width=0.9\columnwidth]{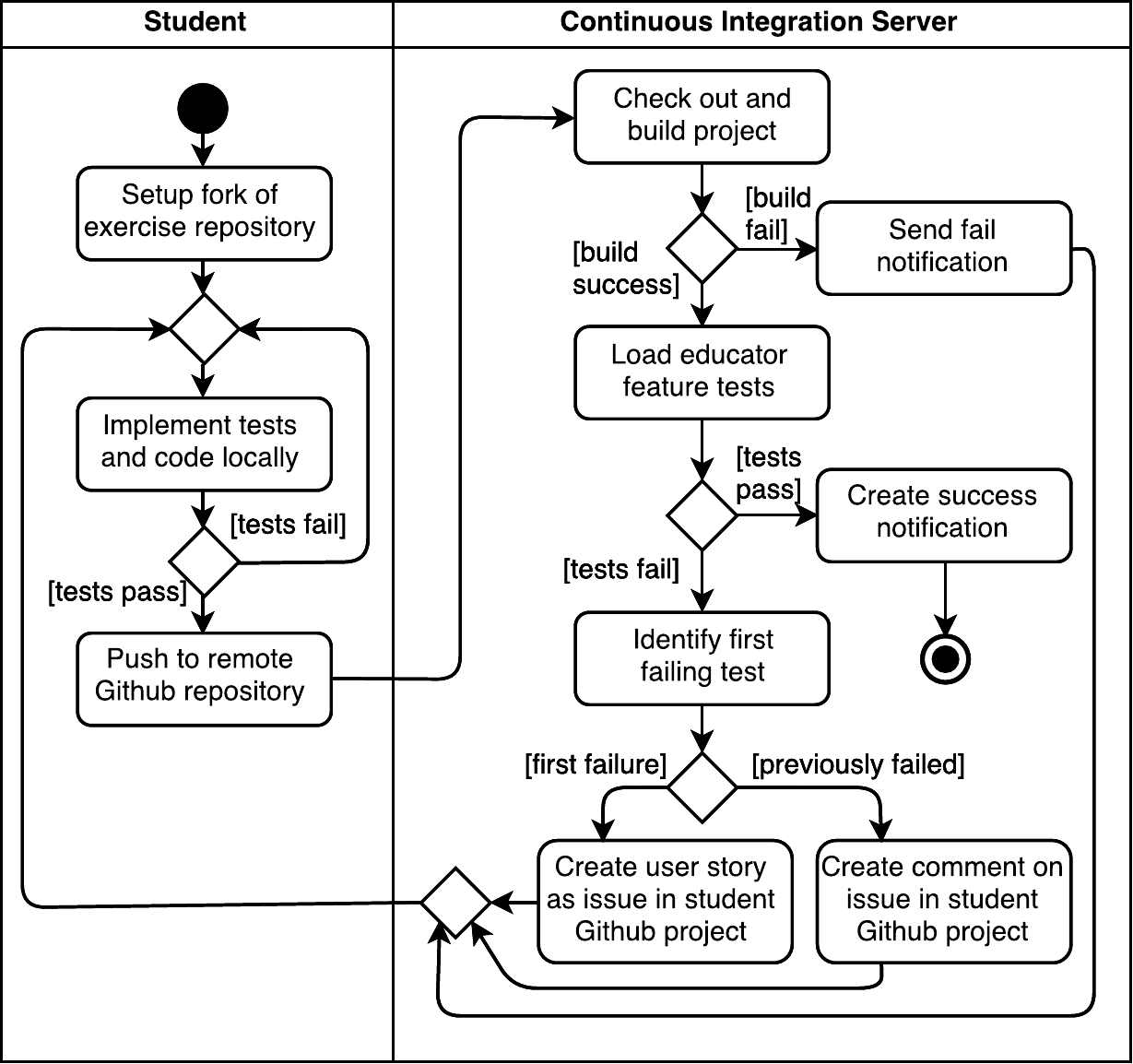}
    \caption{The workflow in a Prof. CI exercise}
    \label{fig:activitydiagram}
\end{figure}

The central piece of a Prof. CI exercise is a continuous integration server which provides tasks to each participant and evaluates their progress.
\Cref{fig:activitydiagram} shows the interaction between participants and the CI server.

Initially, participants receive a link to a public exercise repository.
The repository contains an almost empty project template, a readme file, and the CI scripts that run the validation logic.
The readme contains all the information students need to get started.

To begin the exercise, students fork the exercise repository, set up the \ci\ service for their fork, and clone it to their local machine.

To help students to get started, the project already contains an easily fixable failing test.
For example, in our exercise a heading on the application's front page had to be changed.
When participants push the change, the first CI build is triggered.

Every time the CI server builds a project, it will first conduct a normal build, which includes running all tests written by the students.
If an error occurs, the build will fail and the student receives a notification about the failure via the normal mechanisms of the \ci\ software.

If the build using the participant's test cases was successful, the CI server will run the evaluation script.
This script loads files from a hidden repository containing additional feature tests.
Then, these tests are executed in order and the first failing test will be used to create the next task of the exercise to be solved.

\begin{figure}[t]
    \centering
    \includegraphics[width=\columnwidth]{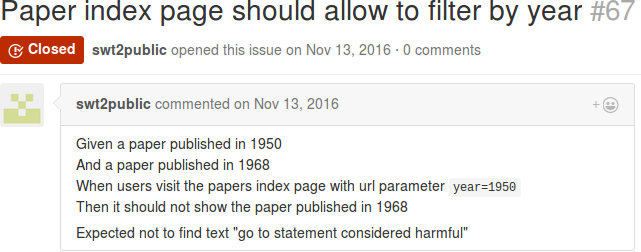}
    \caption{Screenshot of a GitHub issue, opened by Prof. CI in a participant's repository.}
    \label{fig:issue}
\end{figure}

To notify students of the next task, we use the issue tracker of their GitHub repository.
The script will either create a new issue with the test's name as title, or comment on an existing issue of the same name if it is still open.
\Cref{fig:issue} shows an example ticket created by Prof. CI.

The message of the issue or comment always consists of the test's description and error message.
The test description expresses the test logic in natural language and follows the "Given-When-Then" pattern.
It may also contain additional hints such as links to relevant online resources or short code snippets.
The error message of the failed assertion is included to provide a more detailed explanation of why the feature was rejected.
In the exercise introduction, students are instructed to always document the missing feature with a failing test before starting to implement the solution and to prove this by committing the test first.
When all feature tests have passed, students receive a final notification that they have completed the exercise.

In our setup, we did not employ measures to prevent cheating because the exercise was not graded.
However, while no software solution can prevent students sharing their solutions, using private repositories (e.g., with GitHub's classroom feature) can keep students from secretly looking at other students' code.
Furthermore, a nosy student could run the \ci\ script locally to gain access to the secret task repository.
This can be prevented by storing the URL of the task repository in an environment variable on the \ci\ server or by using a \ci\ server that allows running custom code that is not stored in the respective repository.
Finally, while we randomly checked if \tdd\ was used, we suspect that not every student wrote tests first all the time.
For a complete picture, tools can be used to automatically measure how thoroughly students apply the methods of \tdd~\cite{Kou2006Zorro}.

\subsection{Reporting}
\label{sec:reporting}
When relying on a large degree of automation in exercises, it is vital to ensure that problems are detected quickly.
While our approach reduces the direct involvement of educators compared to face-to-face lectures, it also frees up time to analyze and compare course participants based on concrete data.
An analysis of such data is only viable due to automated data collection; manually collecting progress data is infeasible in large user groups.
Our approach of leveraging existing CI tools allows automatically extracting exercise progress data without relying on instrumenting development tools~\cite{johnson2007automated}
or introducing overhead through documentation duties. %instrumenting???
This analysis is advantageous as it gives educators insights into which students are in need of additional coaching or which exact parts of the exercise should be improved in the next iteration.

\subsubsection{Employing GitHub Reports}
\begin{figure}[t]
    \centering
    \includegraphics[width=\columnwidth]{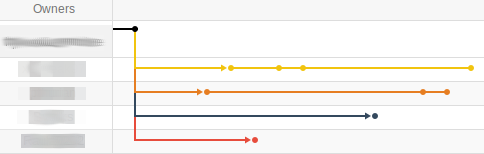}
    \caption{Extract of the GitHub network graph showing commits by course participants, as well as links to their forked repositories.}
    \label{fig:forks}
\end{figure}

\begin{figure}[t]
    \centering
    \includegraphics[width=\columnwidth]{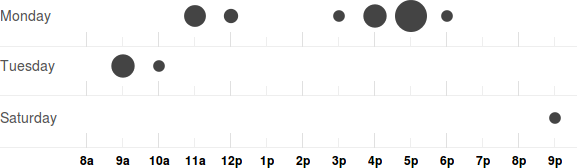}
    \caption{Extract of a course participant's \emph{punch card} visualization of commit times for the last week of the exercise.}
    \label{fig:punch_card}
\end{figure}

GitHub, the platform used for code hosting, already provides tools to analyse and visualize repositories.
In particular, an overview of participants, their forked repositories and their commits is available, cf. \cref{fig:forks}.
This view allows identifying participants who have not committed recently and may be in need of assistance.
Furthermore, more detailed analyses such as \emph{punch cards}, cf. \cref{fig:punch_card}, showing at which times participants committed to their repositories, are available.

\subsubsection{Participant Progress}
\begin{figure}[t]
    \centering
    \includegraphics[width=\columnwidth]{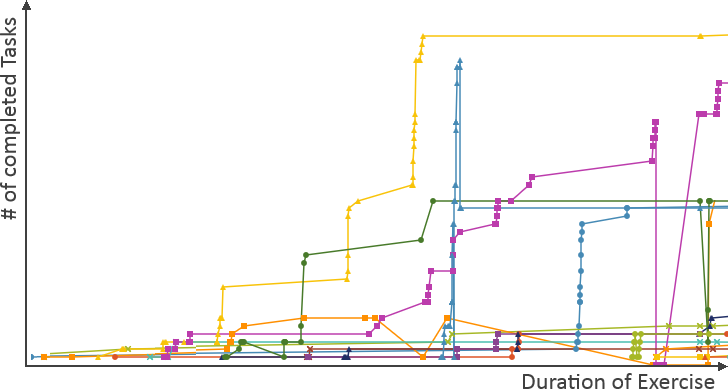}
    \caption{Progress of exercise participants. Every line represents the amount of completed tasks.}
    \label{fig:progress-chart}
\end{figure}

In addition to the analysis available through GitHub directly, we implemented custom visualizations of participant progress.
After a course participant pushes a correctly solved task and moves on to the next task, the user identifier as well as the score, i.e. the number of the task that was solved, and a timestamp are transmitted to a reporting system.
This data is rendered as a live chart of student progress for educators (cf.~\cref{fig:progress-chart}).
It helps to identify those students who might have problems solving a particular task and allows educators to give targeted support, e.g. pointing them to those participants who have already solved the task and might be able to offer assistance.
Furthermore, it provides a way to gauge whether a participant's support request is due to problems with their setup or with the task itself, by visualizing how many other participants have already solved that task.
A decreasing line, i.e. a drop in the score, signifies that the student broke the functionality of an earlier task, which now again fails.

\subsubsection{Time-Per-Task Analysis}
An indicator of both the general difficulty of a task as well as how much effort was required by participants is the time taken to solve tasks.
For every participant of the exercise, the time taken to complete every task is recorded.
This is based on the submitted timestamps of task hand-out and task completion by the CI script, that was run at every build.
Participants that did not trigger a CI build for an hour were considered inactive until the next build occurred.
For the first commit after a break, we assumed 15 minutes of preceding work.
The matrix of time taken per task is presented in \cref{table:timePerTask} for the 2016 exercise instalment.
Blank cells represent tasks that were already already implemented by participants before they were explicitly tasked to do so, i.e. students anticipated the next requested feature.

\begin{table*}[t]
    \caption{Extract of time taken by exercise participants for completing tasks.}
    \label{table:timePerTask}
    \vspace{0.3cm}
    \centering
    \includegraphics[width=\textwidth,trim={2.3cm 17.7cm 3cm 2.5cm},clip]{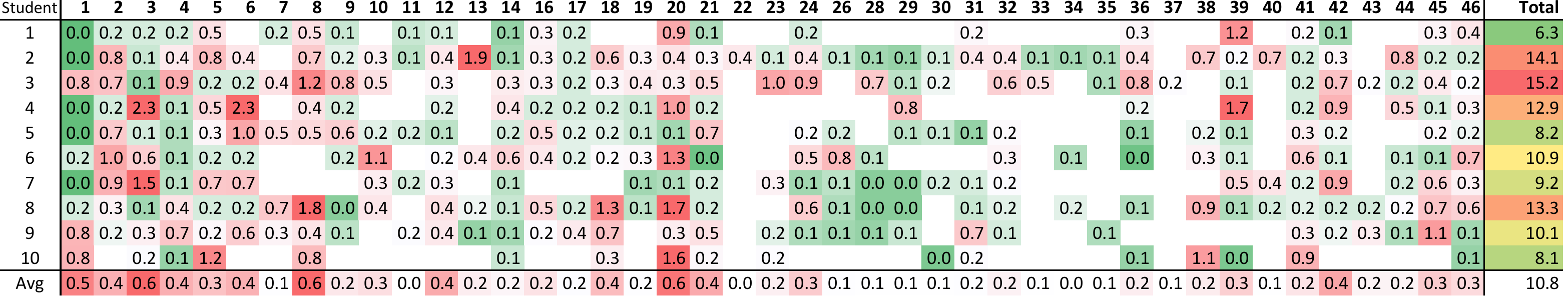}
\end{table*}

This table allows selecting those tasks that took the largest amount of cumulative time over all participants, i.e. those tasks that students had most difficulties with.
Difficult tasks can then be assessed in terms of potential issues, e.g. by collecting targeted feedback or reviewing participants' repositories for solution strategies for this particular task.
If an issue is found that makes the task more time consuming than envisioned, it can be fixed in the next iteration or even while the exercise is running.
For example, task no. 20 contained a test for an HTML element with a default identifier, which failed for students who customized this element.
The task was fixed by using a more general matcher.
Furthermore, the time visualization allows identifying participants who took much longer than others to complete a specific task.
This can be helpful to find those issues that could lead to student frustration and/or quitting~\cite{Mackness2010}.

\section{Evaluation}
\label{sec:evaluation}

We conducted a three week Prof. CI exercise at the beginning of an undergraduate capstone course on software engineering in the 2016 winter term.
The purpose of the exercise was to introduce students to Ruby on Rails and \tdd, before they begin a joint software project using Scrum, which is the main focus of the course.

We evaluated our approach using two different methods:
\begin{itemize}
    \item 
    Immediately after the completing the last task, students were asked to fill out a survey about their opinions on the exercise.
    Results showed positive student attitudes towards the Prof. CI exercise. 
    All participants stated they had understood TDD practices and more than half would recommend the exercise to their peers.
    \item We compared the students' development results in the main software project with the previous year, where no Prof. CI exercise was conducted.
    We found a significant increase in both the number of tests written as well as the number of students writing tests.
    Furthermore, students showed greater awareness towards the importance of code and test quality in the early stages of the project.
\end{itemize}

\begin{figure}[t]
    \centering
    \includegraphics[width=\columnwidth]{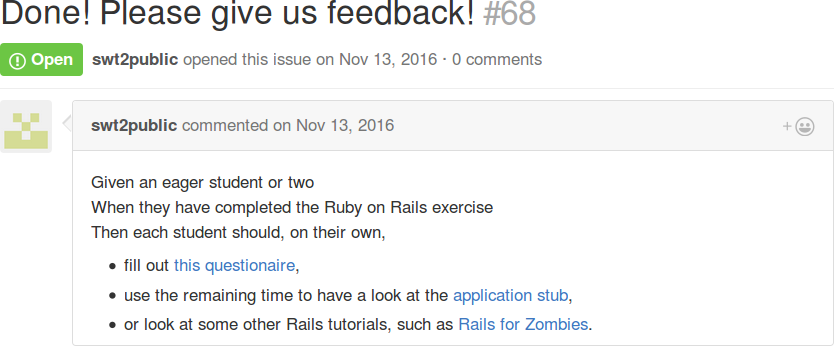}
    \caption{Final ticket given to students upon completing the Prof. CI exercise. Participants are asked to fill out a survey gathering their feedback.}
    \label{fig:final-task}
\end{figure}

\subsection{Surveys}
After completing the final task, students received a last ticket, that provided information on the subsequent development project as well as a link to a survey collecting their opinions on the exercise (cf.~\cref{fig:final-task}).
As such, the survey could be integrated into the flow of the exercise.

\subsubsection{Questionnaire}

The main goal of the survey was to gather the opinions and attitudes of students about two key aspects:
\begin{itemize}
    \item Impressions of the specifics of the Prof. CI exercise, i.e. learning TDD and Ruby on Rails by iteratively solving GitHub tickets.
    \item General feedback on the exercise, its success and improvement possibilities.
\end{itemize}

The questionnaire, featuring twelve questions, is shown in \cref{table:questionnaire}.
Two of the questions (\emph{6a} and \emph{11}) were answerable with free text, the rest could be answered using a 5-point Likert scale (``Strong no'', ``No'', ``Neutral'', ``Yes'', ``Strong yes'').
Overall, 26 students took part in the survey.
The survey was pseudonymous, as students could optionally state their GitHub user name.

\subsubsection{Discussion of Likert-Scale Questions}
\begin{figure}[t]
\centering
    \includegraphics[width=\linewidth,trim={23mm 37mm 24mm 38mm},clip]{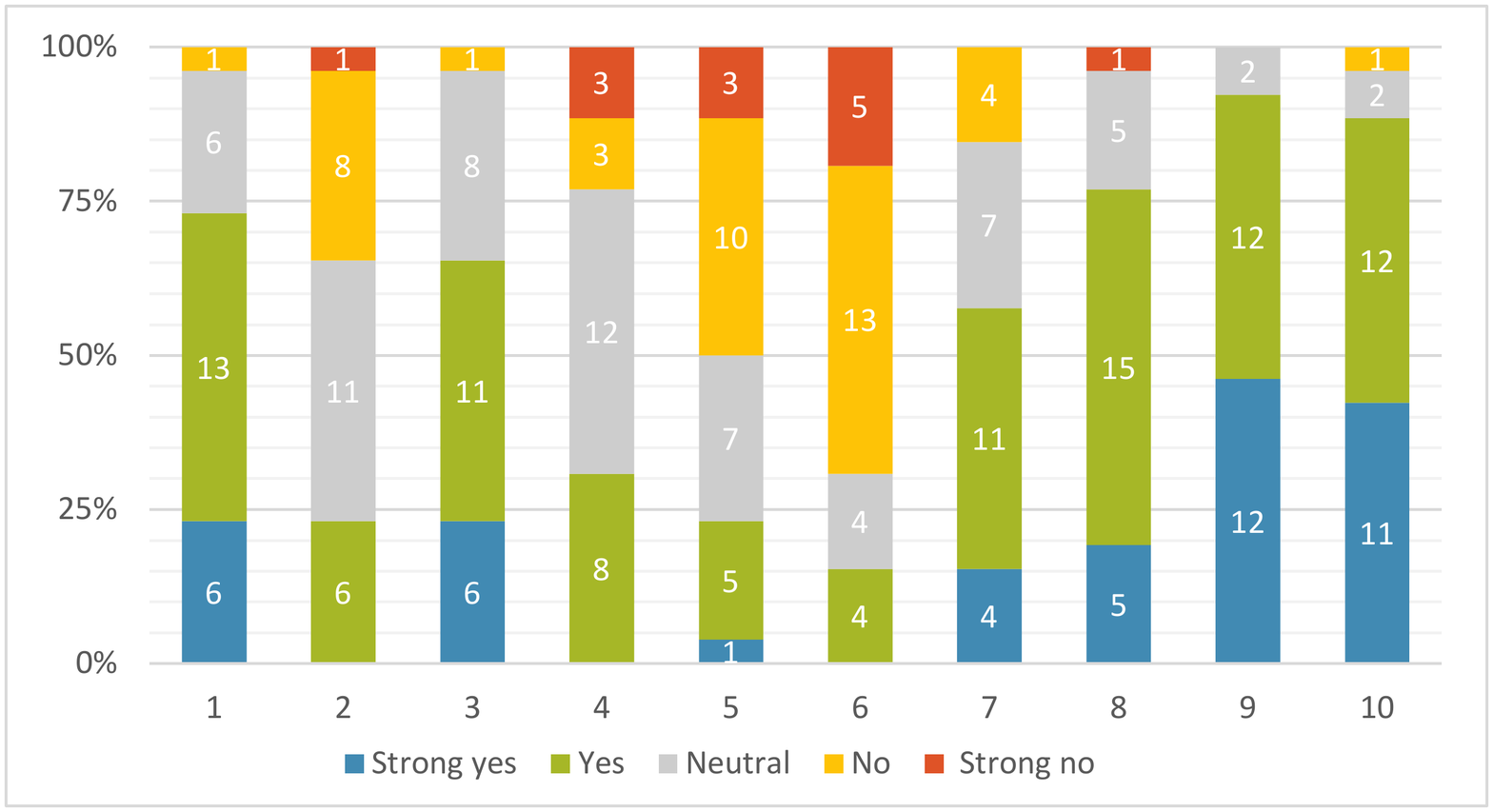}
\caption{Summarized answers of participants to questions 1-10 of the survey.}
\label{fig:survey-answers}
\end{figure}

The answers to the Likert-scale questions of the survey are summarized in \cref{fig:survey-answers}.

One of the major premises of Prof. CI is to require students to write their own test cases, in order to ensure that \tdd\ practices are learnt and applied.
The majority of students (73\%) agreed with this premise, stating that writing their own tests (as opposed to being given finished test cases) helped them learn (\#1).
The vast majority of students (92\%) self-reported that they had understood \tdd\ (\#9) with only positive answers and the highest amount of \emph{strong yes} answers.
The idea of teaching \tdd\ at the same time as a new programming language did not overwhelm students, according to survey answers.
77\% of survey participants stated that the exercise had challenged them (\#8).
However, answers to this question also showed the largest agreement on the \emph{yes} answer, with few extreme answers and only a single \emph{strong no}.
While the practice of using GitHub tickets to communicate new tasks and requirements was generally accepted---65\% of participants stated they were satisfied (\#3)--- issues were also pointed out.
Only 23\% of participants considered the requirements they were given as easy to understand (\#2).
Furthermore, 50\% and 65\% of students disagreed with the statements that they always knew what to do next (\#5) and that they were never unable to continue working (\#6), respectively.
These represent areas of improvement for future exercise instalments.
Even with these shortcomings, 58\% of students stated that they would recommend the Prof. CI exercise to peers (\#7), with no \emph{strong no} answers.
The question with the highest amount of \emph{neutral} answers concerned the question of whether in the next iteration of the exercise all requirements should be available at the start.
With no clear consensus amongst students on this topic, this is an interesting discussion for the future.
While the primary goals of teaching \tdd\ and getting students to write their own tests were fulfilled, based on student answers, there were multiple specific problems in implementation. Furthermore, the Prof. CI exercise also introduced students to \ci\ services. 88\% of participating students answered that they would consider using CI for own projects in the future (\#10).

\begin{table}[t]
    \caption{Feedback survey. Questions \emph{6a} and \emph{11} allowed free text, others used a 5-point Likert scale.}
    \label{table:questionnaire}
    
    \renewcommand{\arraystretch}{1.2}
    \centering
    
    \vspace{0.3cm}
    \textbf{Methodology}
    \vspace{0.05cm}
    
    \begin{tabularx}{\columnwidth}{lX}
        \textbf{\#} & \textbf{Question} \\ \hline
        \textit{1} & Writing my own tests helped me learn. \\
        \textit{2} & The requirements in tickets were easy to understand. \\
        \textit{3} & Working with GitHub tickets that contained new requirements works well.\\
        \textit{4} & In the next iteration of the course all requirements should not be completely available at the start. \\
        \textit{5} & I always knew what to do next. \\
        \textit{6} & At no point was I unable to continue working. \\
        \textit{6a} & If yes, why? What did you do in the meantime? \\
    \end{tabularx}
    
    \vspace{0.3cm}
    \textbf{General Feedback}
    \vspace{0.1cm}
    
    \begin{tabularx}{\columnwidth}{lX}
        %\hline
        \textbf{\#} & \textbf{Question} \\ \hline
        \textit{7} & I would recommend this course to other students. \\ %\hline
        \textit{8} & The exercise challenged me. \\ %\hline
        \textit{9} & I have the feeling of having understood TDD. \\ %\hline
        \textit{10} & I would consider using CI for my own projects. \\ %\hline
        \textit{11} & What would you change in the next course iteration? \\ %\hline
    \end{tabularx}
\end{table}

\subsubsection{Discussion of Free Text Questions}

One of the major drawbacks of using CI services for exercises is the down time that is introduced when waiting for the results of CI builds as well only communicating requirements through short user stories and acceptance criteria in GitHub tickets.
A free text question (\#6a) was introduced to collect student approaches for coping with getting stuck or not being able to continue working.
The gathered answers were tagged with the most common corresponding topics amongst answers.
Of the collected 21 answers, 9 were considered invalid as they did not specify coping methods.
In the remainder, the most common answer (5 mentions) was to use a search engine (specifically Google) to research the problem.
Four participants mentioned asking others for advice, while the strategies of simply waiting, or reading the documentation directly were each only mentioned once.
While all repositories of students were openly accessible, only a single student mentioned checking another participant's repository for the next work items ahead of time.
As the exercise was not graded there was no strong incentive for circumventing the exercise process.

The other free text question (\#11) sought to gather feedback on what students would change in the next iteration of the exercise.
The same tagging procedure as in the previous question was applied.
In the 19 answers that were collected, the most common topic (9 mentions) concerned more instruction regarding test writing.
Other answers mentioned more specific issues (4), an overview of progress through the exercise (3) as well as a better explanation of how the system worked (2).
One student requested a shorter exercise, another requested a mock-up of the functionality to be developed.
These answers strengthen the case for a clearer focus on practicing writing effective tests in university software engineering courses~\cite{janzen2008test}.

Overall, while students reported that there were a number of issues with implementation, the Prof. CI exercise challenged them and the goal of teaching an understanding of Ruby on Rails and \TDD was successful. At the same time, the benefits of using CI services could be demonstrated to students in a hands-on manner.

\subsection{Impact on Subsequent Software Projects}

As the purpose of the Prof. CI exercise was to prepare students for a larger software project, we also evaluated the impact on students' subsequent behavior.

In every iteration of the project, we strongly recommend test-driven development and emphasize the importance of a well-\linebreak maintained test suite.
However, in past years most students started writing tests only late into the project, often only after the lack of a proper test suite had already caused several problems and dissatisfaction with the overall process.
Students reported that despite a lecture on TDD and testing in Rails, as well as a MOOC-based introductory exercise\footnote{In previous years, Codeschool courses, such as \linebreak \url{http://railsforzombies.org/} were employed.}, they did not feel confident in writing tests for their code.

In the previous installment of this lecture, in the 2015 winter term, students were given the same introductory exercise as was used for the Prof. CI one, except that all required feature tests were provided from the start.
During the exercise, students were told to add unit tests for their code and to employ test-driven development.

To compare the students' early project progress, we looked at the results of the first sprint, which was two weeks long in both years.
For each project, we merged all open branches at the last commit before sprint end.

First, we found that in the previous year, in the first two weeks, 44 commits touched the "spec" folder that contains all of the application's tests.
The commits were authored by 10 of the 21 students.
In this year, the "spec" folder was touched by 153 commits over the same period, authored by 24 of 31 students.
This shows that tests were considered more frequently by more students, but it does not paint the full picture.

\begin{figure}[t]
\centering
    \includegraphics[width=\linewidth,trim={27mm 97mm 24mm 111mm},clip]{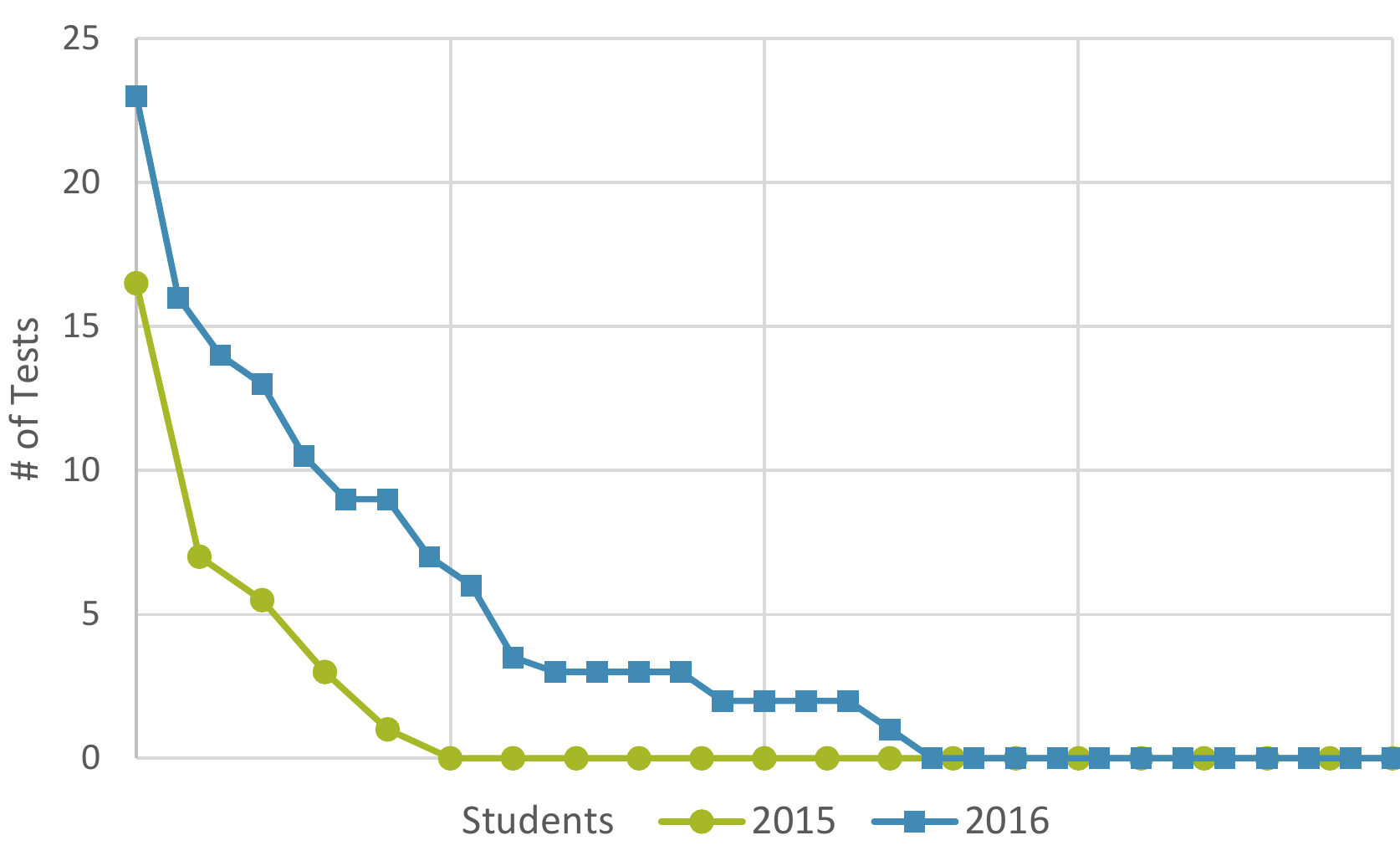}
\caption{Snapshot of the number of manually written or modified tests per student, two weeks into the project. Each data point on the x-axis represents one student, sorted by number of tests.}
\label{fig:testDistribution}
\end{figure}

Because Ruby on Rails automatically generates many tests when application components are created and the previous year's project had a more model classes, it also had a higher total number of tests.
However, of the 377 tests 145 were "pending", i.e., auto-generated test stubs  not completed to form an actual test.
Of this year's 268 tests, none were "pending".
Furthermore, manual changes to the generated code require additional tests.
Thus, we compared the number of manually written tests of both years.
A test was considered "manually written" if it was written from scratch or generated but enriched with meaningful assertions.
\Cref{fig:testDistribution} compares the number of manually written tests per student of both years.
Not only did the average number of tests per student rise from 1.6 to 4.3, but more importantly the ratio of students who wrote at least one test increased by 37 percentage points, from 24 to 61 percent.

Furthermore, the students of the 2016 course set up the \emph{dev} and \emph{master} branches of their project as \emph{protected \linebreak branches}, a GitHub feature that disables force pushes and deletions of these branches and allows a range of custom behavior related to merging.
In particular, it requires \ci\ status checks to pass before allowing merging into these branches.
Students of the previous project did not protect any branches or require \ci\ checks to pass.

Overall, our evaluation shows that not only were students much more confident in writing tests for their code after participating in the Prof. CI exercise, they also placed a greater emphasis on tests as part of their development process.

\subsection{Threats to Validity}

First, we had only 31 students in the experiment and 21 in the control group.
While both groups had the same amount of formal education, individual skill and knowledge can vary, which may lead to statistical artifacts.
Furthermore, all our students were at the end of their Bachelor's studies.
Further studies are needed to evaluate our approach with students at different experience levels.

Second, the exercise of the previous year was not planned to be a control exercise.
As such, available data about the student's performance is incomplete and we could not directly compare students' experiences or the improvement of tool knowledge.

Third, our study is based on only one exercise.
In the exercise, we taught the basic skills necessary to develop in Ruby on Rails with test-driven development.
The effectiveness of our approach for other technology stacks and development methodologies may vary.

Fourth, while we measured a great increase in the number of tests written, we did no formal comparison of test quality except for code coverage.
Likewise, we have no objective comparison of students' proficiency in test-driven development except for their self-reported assessments.

Finally, differences in the subsequent projects' performances might also have been caused by differences in the project's requirements, even though the projects were similar with respect to scope and difficulty.

\section{Conclusion}
\label{sec:conclusion}
We presented \emph{Prof. CI}, an approach for conducting programming exercises, featuring automated incremental task assignment in the form of GitHub issues.
We used this system to introduce students to Ruby on Rails and \tdd\ in an undergraduate capstone course.
Prof. CI features a high degree of automation, as well as possibilities to gather concrete data on student progress and the ability to offer targeted help and continuous improvement of the exercise.
Unlike common browser-based \MOOC\ platforms or more managed setups, students work on their own computers and use \ci\ services to manage tasks and evaluate work.
This allows students to acquire practical skills beyond coding in a new programming language, such as working with a full project and its file tree, running tests locally, working with an issue tracker, and using an IDE and a \vcs.
Contrasting existing approaches that use \ci\ services to evaluate students' work, we built an entire exercise workflow based on \ci.
This allows us to provide software requirements in small incremental tasks, which encouraged testing and the usage of \tdd\ practices.

We found that extra care is required by the exercise authors to ensure the clarity and comprehensibility of each task.
However, our evaluation shows that the approach is effective at teaching complementary skills and increased the students' motivation to write tests for their code in subsequent software projects.

Future work will focus on augmenting the Prof. CI software using the collected feedback, especially more detailed requirements, a clearer explanation of how the system works and an indicator of progress through the exercise.
Furthermore, we will explore how other development practices, apart from \tdd, can be better taught with exercises utilizing \ci\ services and local development tools.


\IEEEpeerreviewmaketitle

\vspace{1cm}

\IEEEtriggeratref{4}
%\bibliographystyle{IEEEtran}
\bibliographystyle{abbrv}
\bibliography{library}  % sigproc.bib is the name of the Bibliography in this case

\begin{thebibliography}{10}

\bibitem{billingsley2013comparison}
W.~Billingsley and J.~Steel.
\newblock A comparison of two iterations of a software studio course based on
  continuous integration.
\newblock In {\em Proceedings of the 18th ACM conference on Innovation and
  technology in computer science education}, pages 213--218. ACM, 2013.

\bibitem{desai2008survey}
C.~Desai, D.~Janzen, and K.~Savage.
\newblock {A survey of evidence for test-driven development in academia}.
\newblock {\em ACM SIGCSE Bulletin}, 40(2):97, 2008.

\bibitem{Fournier2011}
H.~Fournier, R.~Kop, and H.~Sitlia.
\newblock {The value of learning analytics to networked learning on a personal
  learning environment}.
\newblock {\em ACM International Conference Proceeding Series}, pages 104--109,
  2011.

\bibitem{fox2015magic}
A.~Fox, D.~Patterson, S.~Joseph, and P.~McCulloch.
\newblock Magic: Massive automated grading in the cloud.
\newblock In {\em CHANGEE (Facing the challenges of assessing 21st century
  skills in the newly emerging educational ecosystem) workshop at EC-TEL},
  2015.

\bibitem{janzen2008test}
D.~Janzen and H.~Saiedian.
\newblock {Test-driven learning in early programming courses}.
\newblock In {\em ACM SIGCSE Bulletin}, volume~40, page 532. ACM, 2008.

\bibitem{janzen2006test}
D.~S. Janzen, H.~Saiedian, and S.~H. {Janzen D.S.}
\newblock {Test-driven learning: Intrinsic integration of testing into the
  CS/SE curriculum}.
\newblock In {\em Proceedings of the Thirty-Seventh SIGCSE Technical Symposium
  on Computer Science Education}, volume~38, pages 254--258. ACM, 2007.

\bibitem{johnson2007automated}
P.~M. Johnson and H.~Kou.
\newblock {Automated Recognition of Test-Driven Development with Zorro.}
\newblock In {\em AGILE}, volume~7, pages 15--25. Citeseer, 2007.

\bibitem{jones2004test}
C.~G. Jones.
\newblock {Test-driven development goes to school}.
\newblock {\em Journal of Computing Sciences in Colleges}, 20(1):220--231,
  2004.

\bibitem{Kou2006Zorro}
H.~Kou and P.~M. Johnson.
\newblock {\em Automated Recognition of Low-Level Process: A Pilot Validation
  Study of Zorro for Test-Driven Development}, pages 322--333.
\newblock Springer Berlin Heidelberg, Berlin, Heidelberg, 2006.

\bibitem{Mackness2010}
J.~Mackness, S.~Mak, and R.~Williams.
\newblock {The ideals and reality of participating in a MOOC}.
\newblock {\em Networked Learing Conference}, pages 266--275, 2010.

\bibitem{spacco2006helping}
J.~Spacco and W.~Pugh.
\newblock {Helping students appreciate test-driven development (TDD)}.
\newblock In {\em Companion to the 21st ACM SIGPLAN symposium on
  Object-oriented programming systems, languages, and applications}, pages
  907--913. ACM, 2006.

\bibitem{staubitz2015towards}
T.~Staubitz, H.~Klement, J.~Renz, R.~Teusner, and C.~Meinel.
\newblock Towards practical programming exercises and automated assessment in
  massive open online courses.
\newblock In {\em Teaching, Assessment, and Learning for Engineering (TALE),
  2015 IEEE International Conference on}, pages 23--30. IEEE, 2015.

\bibitem{Vihavainen2012}
A.~Vihavainen, M.~Luukkainen, and J.~Kurhila.
\newblock {Multi-faceted Support for MOOC in Programming}.
\newblock {\em Proceedings of the 13th annual conference on Information
  technology education - SIGITE '12}, 68:171--176, 2012.

\end{thebibliography}

\end{document}